\documentclass[sigconf, screen]{acmart}

\AtBeginDocument{%
  \providecommand\BibTeX{{%
    \normalfont B\kern-0.5em{\scshape i\kern-0.25em b}\kern-0.8em\TeX}}}

\usepackage{listings}

\newcommand{\envi}{\ensuremath{\mathit{Env}}}
\newcommand{\cont}{\ensuremath{\mathit{Con}}}
\newcommand{\wpr}{\ensuremath{\mathit{WP}}}

\usepackage[ruled,vlined]{algorithm2e}
\usepackage{microtype}
\usepackage{balance}

\newtheorem{theorem}{Theorem}[section]

\newtheorem{lemma}[theorem]{Lemma}

\setcopyright{acmlicensed}
\acmPrice{15.00}
\acmDOI{10.1145/3468264.3473126}
\acmYear{2021}
\copyrightyear{2021}
\acmSubmissionID{fse21demo-p29-p}
\acmISBN{978-1-4503-8562-6/21/08}
\acmConference[ESEC/FSE '21]{Proceedings of the 29th ACM Joint European Software Engineering Conference and Symposium on the Foundations of Software Engineering}{August 23--28, 2021}{Athens, Greece}
\acmBooktitle{Proceedings of the 29th ACM Joint European Software Engineering Conference and Symposium on the Foundations of Software Engineering (ESEC/FSE '21), August 23--28, 2021, Athens, Greece}

\begin{document}

\title{GenSys: A Scalable Fixed-Point Engine for Maximal Controller
  Synthesis over Infinite State Spaces}

\author{Stanly Samuel}
\email{stanly@iisc.ac.in}
\affiliation{%
  \institution{Indian Institute of Science,}
  \city{Bengaluru}
  \country{India}
}

\author{Deepak D'Souza}
\email{deepakd@iisc.ac.in}
\affiliation{%
  \institution{Indian Institute of Science,}
  \city{Bengaluru}
  \country{India}
}

\author{Raghavan Komondoor}
\email{raghavan@iisc.ac.in}
\affiliation{%
  \institution{Indian Institute of Science,}
  \city{Bengaluru}
  \country{India}
}


\begin{abstract}
The synthesis of maximally-permissive controllers in infinite-state systems has many practical applications. Such controllers directly correspond to maximal winning strategies in logically specified infinite-state two-player games. In this paper, we introduce a tool called GenSys which is a fixed-point engine for computing maximal winning strategies for players in infinite-state safety games. A key feature of GenSys is that it leverages the capabilities of existing off-the-shelf solvers to implement its fixed point engine. GenSys outperforms state-of-the-art tools in this space by a significant margin. Our tool has solved some of the challenging problems in this space, is scalable, and also synthesizes compact controllers. These controllers are comparatively small in size and easier to comprehend. GenSys is freely available for use and is available under an open-source license.
\end{abstract}

\ccsdesc[500]{Theory of computation~Automated reasoning}
\ccsdesc[500]{Theory of computation~Constraint and logic programming}
\ccsdesc[300]{Theory of computation~Logic and verification}

\keywords{reactive synthesis, fixed-points,
  logic, constraint programming 
}

\maketitle

\section{Introduction}

Reactive systems are control programs that continuously interact with
their environment. 
Examples range from cyber physical
systems, robot motion planning systems, wireless sensor networks to bus
arbiters, synchronous and distributed programs, to name a few.
Synthesizing such systems automatically from
temporal specifications 
without human intervention has been a challenge in software engineering for
decades. This problem is of much practical importance, and there are many
approaches in the literature that address it. These approaches can be
classified broadly as ones that address finite-state synthesis~\cite{strix,
  acacia, bosy}, and ones that address infinite-state synthesis~\cite{scots,
  tulip, consynth, simsynth,jsyn}.

While modelling a reactive system, we can view it as a game between two non
co-operating players, with a given winning condition.  The
\emph{controller} is the protagonist player for whom we wish to find a
strategy, such that it can win against any series of moves by the other
player, which is the \emph{environment}.  A play of the game is an infinite
sequence of \emph{steps}, where each step consists of a move by each
player.

The aim of synthesis is to find a ``winning region'' and a winning strategy
for the controller if these exist. A winning region consists of a set of
states from which 
the controller will win if it follows
its strategy. 

In addition to scalability, speed, and size of the
synthesized control program, 
the quality of ``maximal
permissiveness,'' which requires the program to
allow as many of its moves as possible while still guaranteeing
a win, has also gained importance in recent applications.
A \emph{maximal} winning region is one that contains
all other winning regions.
For instance, a maximally permissive program could be used as a ``shield'' for a neural
network based controller \cite{ZhuXMJ19}, and a maximal control program
would serve as the ideal shield.  Another practical application of reactive synthesis for software engineering is in the domain of model based fuzz testing and has been explored in \cite{fuzz}.

In this paper we introduce our tool GenSys, which performs efficient
synthesis of 
\emph{maximal} control programs, for infinite-state systems.
Gensys uses a standard fixpoint computation \cite{wolfgang}
to compute a maximal controller, and does so by leveraging the tactics
provided by off-the-shelf solvers like Z3 \cite{z3}.
Our approach is guaranteed to find a maximal
winning region and a winning strategy for any given game whenever the
approach terminates.

GenSys is available on
GitHub\footnote{\url{https://github.com/stanlysamuel/gensys}}.

\section{Motivating Example}
A classic example of a game with infinite states is that of
Cinderella-Stepmother \cite{cinderella1, cinderella2}. This has been
considered a challenging problem for automated synthesis. The game is
practically motivated by the minimum backlog problem~\cite{mbp}, which is an
online problem in the domain of wireless sensor networks.

The game consists of five buckets with a fixed capacity of $C$ units
each, arranged in a circular way. 
The
two players of the game are Cinderella, who is the controller, and the
Stepmother, who is the environment. 
In each step, Cinderella is allowed to empty any two adjacent
buckets, and then the Stepmother tops up the buckets by arbitrarily
partitioning one fresh unit of liquid across the five buckets. Cinderella
wins if throughout the play none of the buckets overflow; otherwise the
Stepmother wins.

The winning region for Cinderella 
in the Cinderella-StepMother game with bucket capacity
three units comprises states where three consecutive buckets have at
most two  units each, with the sum of the first and third of these buckets
being at most 3 (see
Table~\ref{table: c3}).

We will use this game as a running example to illustrate the
components of the tool.

\section{Tool Design}

\begin{figure}
  \centering
  \includegraphics[width=\linewidth]{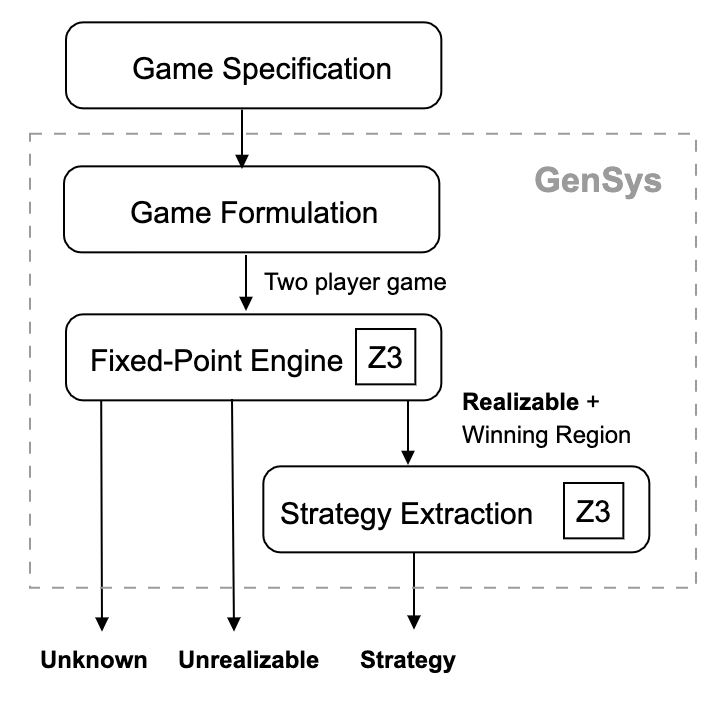}
  \caption{GenSys  Tool Architecture}
  \label{fig:tool}
  \Description{Contains Game Formulation, Fixed-Point Engine and Controller Extraction module}
\end{figure}

GenSys allows users to model a reactive game, to provide a winning
condition, and to check automatically if a strategy can be synthesized for
the controller to win the game.  Figure~\ref{fig:tool} describes the
overall architecture of GenSys. We describe the main components of the
tool below.

\subsection{Game Specification}
\label{sec:input}

The \emph{game specification} is given as input by the user, and consists
of four parts: the state space, environment moves, controller moves, and
the \emph{winning condition}. A sample game specification is depicted in
Figure~\ref{fig:code}, corresponding to the Cinderella-Stepmother game. The
game specification needs to be Python code, and needs to make use of certain
API features provided by GenSys. In  Figure~\ref{fig:code} we have
used three buckets for brevity; in our evaluation
we use five buckets as that is the standard configuration used in the
literature.  


\begin{figure}
  \input{code.tex}
  \caption{Cinderella Game Specification in GenSys}
  \label{fig:code}
\end{figure}

\subsubsection*{State space:}

Every game consists of a \emph{state space}, where a state consists of a
valuation for a set of \emph{variables}. In the example in
Figure~\ref{fig:code}, the variables are named b1, b2, and b3. Intuitively,
the values of these variables represent the amount of liquid in each bucket
currently.  GenSys follows the convention that a variable name of the
form ``\textit{var}\_''  represents the ``post'' value of
``\textit{var}''  after a move.

\subsubsection*{Environment move:}
Lines~6--7 define the state-update permitted to the environment (which
would be the StepMother in the example) in each of its moves. In
Figure~\ref{fig:code}, this portion indicates that the StepMother can add a
total of one unit of liquid across all three buckets.
Semantically, the environment moves can be encoded as a binary relation
$\envi(s,s')$ on states. 

\subsubsection*{Controller move:}
This portion defines the state-update permitted to the controller (which
would be Cinderella in the example) in each of its moves.  Lines~10--19 in
the code in Figure~\ref{fig:code} indicate that the controller has three
alternate options in any of its moves.   `move1'
corresponds to emptying buckets b1 and b2, and so on.  Semantically,
the controller moves can be encoded as a binary relation $\cont(s,s')$ on
states.  In Figure~\ref{fig:code}, $\cont(s,s')$ is a disjunction of each controller move in the Python list \textit{controller\_moves}.



\subsubsection*{Safe Set:} We support \emph{safety} winning
conditions as of now in GenSys.
A safety winning condition is specified by a set of ``safe'' states in
which the controller must forever keep the play in, to win the play.
In Lines 24--25, the safe set of states 
is given by the condition that each bucket's 
content must be at most the bucket capacity
$C$, which is a command-line parameter to the tool. In other words,
there should be no 
overflows. Semantically, the safe set 
is a predicate $G(s)$ on states.

To solve the safety game, the user should call the \textit{safety\_fixedpoint} function which implements the fixed-point procedure for this winning condition.  This function takes as input moves of both players and the safe set and returns a strategy for the controller, if it exists.  More details regarding the procedure is explained in Sections \ref{ssec:game-formulation}, \ref{ssec:fixedp-engine} and \ref{ssec:strategy-extraction} respectively.

In this prototype version,  there is no formal specification language and the game specification needs to be python functions in a specific format,  as shown in Fig \ref{fig:code}.  More details can be found on our tool page\footnote{\url{https://github.com/stanlysamuel/gensys}}.  Support for initial variables is not incorporated but is a trivial extension.


\subsection{Game Formulation}
\label{ssec:game-formulation}
From the given game specification, this module of our tool formulates one
step of the game.  This step is represented as the following equation:


$$
\begin{array}{lcl}
\wpr(X)& \equiv & \exists s' (Con(s,s') \ \wedge G(s') \ \wedge\\
&& \ \forall s'' (
Env(s', s'') \implies  X(s''))).
\end{array}
$$

A step consists of a move of the controller followed by a move of the
environment.  The formula above has the state variable $s$ as the free
variable. The solution to this formula is the set of states starting from
which the controller has a move such that if the environment subsequently
makes a move,  
the controller's move ends in a state that satisfies the given winning
condition $G$, and the environment's move ends in a state that is in a
given set of states $X$. The formula above resembles the weakest
pre-condition computation in programming languages.  Note that the
controller makes the first move
\footnote{We also support the scenario where
  the environment plays first but this is beyond the scope of this paper.}.


\subsection{Fixed-Point Engine}
\label{ssec:fixedp-engine}
The \emph{winning region} of the game is the greatest solution to the equation in Section \ref{ssec:game-formulation} and can be represented by the greatest fixed-point expression:
\begin{displaymath}
\nu X. \  (\wpr(X)  \wedge  G)
\end{displaymath}

It should be noted that for soundness, we require that $X$ be initialized to $G$ as opposed to $True$ in the standard gfp computation.

The winning region represents the set of states starting from which the
controller has a way to ensure that only states that satisfy the winning
condition $G$ are visited across any infinite series of steps. Our tool
computes the solution to the fixed-point equation above using an iterative
process (which we describe later in the  paper). 

Our formulation above resembles similar classical formulations for finite
state systems~\cite{pnuelimaler,wolfgang}. Those algorithms were guaranteed
to terminate due to the finiteness of the state space.  This is not true in
the case of an infinite state space.  Thus, it is possible our approach
will not terminate for certain systems. In Figure~\ref{fig:tool}, this
possibility is marked with the ``Unknown'' output.
Thus,  we are
\emph{incomplete} but \emph{sound}. 
We note that due to the uncomputable nature of the problem \cite{simsynth} there
cannot exist a terminating procedure for the problem.  However, we have empirically observed that if we bound the variables in $G(s)$,  the procedure terminates.  For example,  for the cinderella specification in Fig \ref{fig:code},  if we use the constraint $ \bigvee_{i = 1}^{3} b_i <= C$ for $G(s)$,  the procedure does not terminate.

\emph{Maximality:} If the procedure terminates,  the winning region is maximal i.e.,  it contains the exact set of states from where the controller can win.  For the proof sketch,  assume that the region is not maximal.  Then there exists a state which was missed or added to the exact winning region.  This is not possible due to the fact that at every step,  the formulation in Section \ref{ssec:game-formulation} computes the weakest set of states for the controller to stay in the safe region,  against any move of the environment. The detailed proof can be found in Section \ref{sec-appendix}.



\subsection{Strategy Extraction}
\label{ssec:strategy-extraction}
The game is said to be winnable for the controller, or a winning strategy
for the controller is said to be \emph{realizable}, if the winning region
(computed above)  is non-empty. 

From the winning region, the strategy can be emitted using a simple logical
computation.  The strategy is a mapping from subsets of the winning region
to specific alternative moves for the controller as given in the game
specification, such that every state in the winning region is present in at
least one subset, and such that upon taking the suggested move from any
state in a subset the successor state is guaranteed to be within the
winning region.

In the Cinderella-StepMother game, when there are five buckets and the
bucket size $C$ is 3, the strategy that gets synthesized is shown in
Table~\ref{table: c3}.

\begin{table}
\caption{Strategy Synthesized by GenSys for the Cindrella game with bucket size $3$}\label{table: c3}
\begin{tabular}{cc}
\toprule
Condition & Move \\
\midrule
 $0\leq b_1, b_2 \leq 3 \, \wedge\, 0 \leq b_3, b_4,
b_5 \leq2 \, \wedge \, b_3 + b_5 \leq 3$ &  $b_1\_, b_2\_ = 0$  \\
$0\leq b_2, b_3 \leq 3 \, \wedge\, 0 \leq b_4, b_5, b_1\leq 2 \, \wedge
\, b_4 + b_1 \leq 3$ &  $b_2\_,  b_3\_ = 0$  \\
$0 \leq b_3, b_4 \leq 3 \, \wedge \, 0 \leq b_5, b_1, b_2 \leq 2\, \wedge
\, b_5 + b_2 \leq 3$ &  $b_3\_,  b_4\_ = 0$  \\
$0 \leq b_4, b_5 \leq 3 \, \wedge\, 0 \leq b_1, b_2, b_3 \leq 2 \, \wedge
\, b_1 + b_3 \leq 3$ &  $b_4\_, b_5\_ = 0$ \\
$0 \leq b_5, b_1 \leq 3 \, \wedge \, 0 \leq b_2, b_3, b_4 \leq 2 \, \wedge
\, b_2 + b_4 \leq 3$ &  $b_5\_, b_1\_ = 0$ \\
\bottomrule
\end{tabular}
\end{table}

It is interesting to note that  a sound and readable strategy 
has been synthesized automatically,  without any human in the loop.

\section{Implementation Details}
GenSys is currently in a prototype implementation stage, and serves as a
proof of concept for the experimental evaluation that follows.  The current
version is 0.1.0. Currently GenSys supports safety winning conditions;
immediate future work plans include adding support for other types of
temporal winning conditions.

GenSys is implemented in Python, and depends on the Z3 theorem
prover~\cite{z3} from Microsoft Research. GenSys has a main loop, in which it
iteratively solves for the fixed-point equation in
Section~\ref{ssec:fixedp-engine}. It first starts with an
over-approximation $X = G$, where $G$ is the given safe set, and computes
using Z3 a formula that encodes $\wpr(X)$. It then makes $X$ refer to the
formula just computed, re-computes $\wpr(X)$ again, and so on iteratively,
until the formulas denoted by $X$ do not change across iterations.  This procedure is described in Section \ref{sec-appendix}.

The iterative process above, if carried out naively,  can quickly
result in very large formulas. To mitigate this issue, we make use of Z3's
quantifier elimination tactics. Z3 provides many such tactics; our studies
showed that the `qe2'~\cite{qe2} strategy showed the best results. We
believe the quantifer elimination power of Z3 is one of the main reasons
for the higher scalability of our approach over other existing approaches.

\section{Experimental Results}

To evaluate our tool GenSys, we consider the benchmark suite from the paper
of Beyene et al.~\cite{consynth}, which introduces the Cinderella game as
well as some program repair examples.  We also consider the robot motion
planning examples over an infinite state space introduced by Neider et
al.~\cite{automatonlearning}.

The primary baseline tool for our comparative evaluation is
JSyn-VG~\cite{jsyn}, whose approach is closely related to ours.  Their
approach also uses a weakest-precondition like formulation and an iterative
approach to compute a fix-point solution. However, their approach uses a
``forall-there-exists'' formulation of a single step, in contrast to the
``there-exists-forall'' formulation that we adopt (see the $\wpr$
formulation in Section~\ref{ssec:game-formulation}). Also, their tool uses
a dedicated solver called AE-VAL~\cite{aeval1, aeval2}, whereas GenSys uses
the standard solver Z3. 

 We used the latest version of the JSyn-VG, which is available within the
JKind model checker
(\url{https://github.com/andrewkatis/jkind-1/releases/tag/1.8}), for our
comparison.  

To serve as secondary baselines, we compare our tool with several other tools on the same set
of benchmarks as mentioned above. These tools include SimSynth~\cite{simsynth}
and ConSynth~\cite{consynth}, which are based on logic-based synthesis,
just like GenSys and JSyn-VG. We also consider the tool
DT-Synth~\cite{dtsynth}, which is based on decision tree learning, and the
tools SAT-Synth and RPI-Synth, which are based on automata based
learning~\cite{automatonlearning}. The numbers we show for SimSynth
and ConSynth are reproduced from ~\cite{simsynth} and ~\cite{dtsynth} respectively, while the numbers for all
other tools mentioned above were obtained by us using runs on a machine with an Intel i5-6400 processor and 8 GB RAM.  \footnote{We were unable to build SimSynth from source due to the dependency on a very specific version of OCaml.  We were unable to get access to ConSynth even after mailing the authors.  Thus,  we used the numbers for ConSynth from the DT-Synth ~\cite{dtsynth} paper which is the latest paper that evaluates ConSynth. They also describe the difficulty in reproducing the original ConSynth results.  We  expect the ConSynth results that we have reproduced from the other paper \cite{dtsynth} to be accurate,  as the numbers for  the other tools given in that paper match the numbers we obtained when we ran those tools.}
Results for the Cinderella game are not available
from the learning-based approaches (i.e.,  they time out after 900 seconds).  SimSynth
results are available only for Cinderella among the benchmarks we consider. 


\begin{table}
  \caption{Running times for the Cinderella game for various values of
    bucket size $C$.  "-" indicates unavailability of data, while ">$x$m" denotes a timeout after $x$ minutes.  R denotes Realizable and U denotes Unrealizable.}
\label{table: cinderella}
\begin{tabular}{lrrrrrrrr} 
\toprule
$C$ & Out &  SimSynth &  ConSynth & JSyn-VG &  \multicolumn{2}{c}{GenSys}\\
\cmidrule(lr){6-7}
& & &  & & Time & Iter \\
\midrule
3.0 & R & 2.2s &  12m45s & 1m26s &0.6s & 3\\
2.5 &R & 53.8s  & \textbf{>15m} &1m19s&0.7s &  3\\
2.0 &R & 68.9s  & - &1m6s&0.6s & 3\\
1.9(20) & U & - &  -&\textbf{>16m}&31.0s & 69\\
1.8 &U & \textbf{>10m}  & - &\textbf{>16m}&0.6s & 5\\
1.6 &U & 1.5s  &  -&\textbf{>16m}&0.4s & 4\\
1.5 &U & 1.4s  &  - &14m34s&0.3s & 4\\
1.4 &U & 0.2s  &  - &17s&0.2s & 3\\
\bottomrule
\end{tabular}
\end{table}

Table~\ref{table: cinderella} contains detailed results for the Cinderella
game, by considering various values for the bucket size $C$. It was
conjectured by the ConSynth tool authors~\cite{consynth} that the range of
bucket sizes between $ \geq 1.5$ and $<2.0$ units is challenging, and that
automated synthesis may not terminate for this range.  They also mention that this problem was posed by Rajeev Alur as a challenge to the software synthesis community.  However, GenSys
terminated with a sound result throughout this range.  In fact, GenSys was able to
scale right upto bucket-size 1.9(20) (i.e., the digit 9 repeated 20 times
after the decimal), whereas the state of the art tools time out much
earlier. The number of iterations for the fixed-point loop to terminate,
i.e., 69, and the time taken to solve, i.e., 31 seconds, affirm that it was
indeed challenging to solve for this bucket size.  This empirically proves
that we can scale to large formula sizes. This is challenging because the formula sizes
keep increasing with every iteration of the fixed-point computation. 

Table~\ref{table: others} shows the results on the other benchmarks. Here
also it is clear that GenSys outperforms the other tools in most
situations.  

SimSynth supports reachability,  which is a dual of safety.  ConSynth supports safety, reachability and general LTL specifications.  The rest of the tools that we consider,  including GenSys,  natively support safety (and its dual,  reachability) winning conditions only.

Regarding maximality,  it should be noted that JSyn-VG is the only tool apart from us that synthesizes a maximal controller.  

\begin{table}
  \caption{Results on remaining benchmarks. Times are in seconds.
    \textbf{>15m} denotes a timeout after 15 minutes. Tool name
    abbreviations: C for ConSynth, J for JSyn-VG, D for DT-Synth, S for
    SAT-Synth, R for RPI-Synth, G for GenSys.}
\label{table: others}
\begin{tabular}{lrrrrrrrrr} 
\toprule
Benchmark &  C & J &
D & S &
R &   G \\
\midrule
Repair-Lock& 2.5&1.5&   0.5 & 0.6 & 0.2 &0.3\\
Box&3.7&0.6& 0.3 & 0.3  & 0.1 &0.3\\
Box Limited &0.4&1.7& 0.1 & 0.4 & 0.5 &0.2\\
Diagonal&1.9&4.0& 2.4 & 1.34 & 0.5 &0.2\\
Evasion&1.5&0.5& 0.2  & 81 & 0.1 &0.7\\
Follow&\textbf{>15m}&1.2& 0.3 & 88.9 & \textbf{>15m} &0.7\\
Solitary Box&0.4&0.9& 0.1 & 0.3  & 0.1 &0.3\\
Square 5x5&\textbf{>15m}&6.5& 2.5 & 0.6 & 0.2 &0.3\\
\bottomrule
\end{tabular}
\end{table}

\section{Future Work}
The scalability of our approach hints at the potential for addressing more
complex winning conditions apart from safety.  It would be interesting to
address synthesis of maximal controllers for $\omega$-regular
specifications, which is a strict superclass of safety, and compare
scalability, synthesis time, and controller size for such properties.

\section{Conclusion}
We have presented the prototype implementation of our tool GenSys.  We
discussed the design of the tool using a motivating example, and
demonstrated scalability of strategy synthesis and the readability of
synthesizied strategies. One of the key takeaways is that with the advances
in SMT algorithms for quantifier elimination and formula simplification, it
is possible to expect scalability for fundamental problems.  Tools such as
ConSynth,  JSyn-VG and SimSynth use external solvers such as E-HSF~\cite{ehsf},  AE-VAL~\cite{aeval1, aeval2}, and SimSat \cite{simsat} respectively, which appear to slow down the synthesis
process.  E-HSF requires templates for skolem relations, while AE-VAL
restricts the game allowing only the environment to play first.  Although SimSynth does not require external templates as a manual input,  it follows a two step process where it first synthesizes a template automatically using SimSat, followed by the final strategy synthesis.  Our
approach does not require an external human in the loop to provide
templates,  does not pose restrictions on the starting player and is a relatively intuitive approach.
Thus, we show an elegant solution that works well in practice.  More information about our approach,  running the tool and reproducing the results can be found on GitHub\footnote{\url{https://github.com/stanlysamuel/gensys}}.



\bibliographystyle{ACM-Reference-Format}
\bibliography{gensys}

\newpage
\section{Appendix}
\label{sec-appendix}

\subsection{Safety Procedure}
\label{safety-algorithm}
Algorithm \ref{proc: safety} computes the greatest solution to the equation in Section \ref{ssec:game-formulation}.

\begin{algorithm}
\SetAlgoLined
\SetKwInOut{Input}{Input}
\SetKwInOut{Output}{Output}
\Input{ Game formulation \wpr, Safe region $G$}
\Output{ Winning region $X$,  if algorithm terminates}
 $X$ := $G$ \;
 $W$ := $\wpr(X) \wedge G$ \;
 \While{$X \nRightarrow W$}{
  $X$ := $W$\;
  $W$ := $\wpr(X) \wedge G$
 }
 return $X$;
 \caption{Safety Procedure}
 \label{proc: safety}
\end{algorithm}


Algorithm \ref{proc: safety} takes the game formulation as input and returns the winning region for the controller,  if it terminates. The winning region is a quantifier free formula in the base theory.  At every iteration, the formula $\wpr(X) \wedge G$ is projected to eliminate quantifiers to return an equivalent quantifier free formula $W$. The projection operation is intrinsic to the Z3 solver.

\subsection{Proof:}
We prove the correctness of the Algorithm \ref{proc: safety} by reasoning over $X$.

\begin{lemma}
\label{lemma}
 At the $i$'th step of Algorithm \ref{proc: safety}, $X_i$ is the exact set of states from where the controller has a strategy to keep the game in G for at least $i$ steps.
\end{lemma}

\emph{Proof:}  We prove this by induction over the valuations of predicate $X$ at every step in Algorithm \ref{proc: safety}.

Base case: $i=0$ and $X_0 = G$.  Trivially, the game stays in $G$ and hence it is the set of states from where the controller has a strategy to keep the game in G for at least $0$ steps. This is also the weakest (and hence exact) set of states as there are no other states from where the controller can win without making a move.

Inductive step:
Assume that the IH holds i.e.,  $X_{i-1}$ is the exact set of states from where the controller has a strategy to keep the game in G for at least $i-1$ steps.

$X_i$ is computed as $X_i := \wpr(X_{i-1}) \wedge G$.  From any state $s \in X_i$, the controller can stay in the safe region and ensure reaching $ X_{i-1}$ in one step ensuring the fact that it can keep the game in $G$ for at least $i$ steps.  Hence, $X_i$ is sound.

\emph{Claim:} $X_i$ is the weakest.

\emph{Proof:} Assume a state $s \notin X_i$ and from where the controller can ensure a win.  This is not possible because $s$ must be a solution to $\wpr \wedge G$.

\begin{theorem}[Soundness]
The predicate $X$ returned by Algorithm \ref{proc: safety} is a winning region for the controller.
\end{theorem}

\emph{Proof:} Let $X_{k+1} = X_k$ for some step $k$ in Algorithm \ref{proc: safety}.
Let $s \in X_{k+1}$.  From Lemma \ref{lemma},  $X_k$ is the exact set of states from where the controller has a strategy to keep the game in G for at least $k$ steps.  Similarly,  the lemma holds for $X_{k+1}$.
Since $ X_{k+1}  = \wpr(X_k)$,  from $s$,  the controller can ensure a move to reach $X_k$ in one step.  Since $X_{k+1} = X_k$, the controller can ensure a move to reach $X_{k+1}$ in one step as well.  As this process can be repeated forever, $X_k$ (and hence,  $X$) is a winning region.

\begin{theorem}[Maximality]
$X$ returned by Algorithm \ref{proc: safety} is the weakest region i.e.,  no state from where controller can win, is missed.
\end{theorem}

\emph{Proof:}
Assume not. Then there exists a state $s \notin X$ from which the controller can keep the game in the safe region for infinite steps.  Let the algorithm terminate at some step $k$.  By Lemma \ref{lemma}, $X_k$ is the exact set of states from where the controller has a strategy to keep the game in G for at least $k$ steps. Infinite steps also include the $k$'th step of the algorithm,  since $k$ is arbitrary. 
Hence $s \in X_k$.  Contradiction.

From the above two theorems,  $X$ is sound and the weakest set of states from where the controller can ensure a move.

\subsection{Strategy Extraction:}

Once the winning region $X$ has been computed,  the strategy for the controller can be extracted in one step.  In this paper,  we assume that the controller is a disjunction of finite number of moves.  Thus,  for $n$ moves:
\begin{displaymath}
Con(s,s') = \bigvee_{i = 1}^{n} Move_i(s,s')
\end{displaymath}
Let
$$
\begin{array}{lcl}
\wpr_i(X)& \equiv & \exists s' (Move_i(s,s') \ \wedge G(s') \ \wedge\\
&& \ \forall s'' (
Env(s', s'') \implies  X(s''))).
\end{array}
$$
Given the winning region $X$, the strategy extraction step computes the condition under which each move of the controller should be played,  as follows:

\begin{displaymath}
Condition_i = \wpr_i(X) \wedge G
\end{displaymath}

For $n$ moves,  the strategy returned is a map from conditions to moves as follows:

\begin{displaymath}
Condition_i \mapsto Move_i,\ \ \ \ \ \   i \in \{ 1 .. .  n\}
\end{displaymath}

\emph{Soundness and Maximality of the synthesized strategy:}
This follows from from the soundness and maximality of the winning region $X$.  The nuance is that the argument now depends on each move $Move_i(s,s')$ of the controller instead of $Con(s,s')$.

%
%
%


\end{document}